\documentclass[conference]{IEEEtran}

\pdfoutput=1

\newcommand{\eas}[2]{\begin{subequations}\begin{eqnarray} #1 \end{eqnarray}{{#2}}\end{subequations} }
\newcommand{\lb}{\left(}
\newcommand{\rb}{\right)}
\newcommand{\lsb}{\left[}
\newcommand{\rsb}{\right]}

\newcommand{\ea}[1]{\begin{eqnarray} #1 \end{eqnarray} }

\newcommand{\labs}{\left|}
\newcommand{\rabs}{\right|}

\newcommand{\al}{\alpha}
\newcommand{\be}{\beta}
\newcommand{\alb}{\bar{\alpha}}

\newcommand{\Ccal}{\mathcal{C}}
\newcommand{\f}[2]{\frac {#1} {#2}}

\newcommand{\p}[1]{\begin{array}{l l l l l l l l l l} #1 \end{array} }

\newcommand{\RR}{\mathbb{R}}

\newcommand{\cov} {\mathrm{Cov}}

\renewcommand{\Re} {\mathrm{Re}}

\renewcommand{\Pr} {\mathop\mathrm{Pr}\nolimits}

\newcommand{\E}{\mathbb{E}}
\newcommand{\reff}[1]{\eqref{#1}}
\newcommand{\defeq}{\triangleq}

\newcommand{\Cc}{\mathcal{C}}

\newcommand{\Nc}{\mathcal{N}}

\newcommand{\zerov}{\boldsymbol{0}}
\newcommand{\sgs}{\sigma^2}

\usepackage{cite}      

\usepackage{graphicx}  

\usepackage{psfrag}    

\usepackage{subfigure} 

\usepackage{url}       

\usepackage{stfloats}  

\usepackage{amsmath}   
\usepackage{array}
 \makeatletter
 \let\NAT@parse\undefined
 \makeatother
\usepackage{times}
\usepackage{epsfig,graphicx,color,pstricks}
\usepackage{amsmath,amssymb,amsbsy,amsthm,amsfonts,latexsym}
\usepackage{bm}
\usepackage{psfrag}
\usepackage[mathscr]{eucal}
\usepackage{cite}
\usepackage[all]{xy}
\usepackage{subfigure}
\usepackage{graphics}

\newtheorem{thm}{Theorem}

\theoremstyle{remark}

\theoremstyle{definition}

\def \gap {3~}

\begin{document}
\title{Capacity to within \gap bits for a class of Gaussian Interference Channels with a Cognitive Relay}
\author{%
\IEEEauthorblockN{Stefano Rini, Daniela Tuninetti and Natasha Devroye}
\IEEEauthorblockA{ECE Department, University of Illinois at Chicago, Chicago, IL 60607, USA,\\
Email: {\tt {srini2, danielat, devroye}@uic.edu} }
}
\maketitle

\maketitle

\begin{abstract}
The InterFerence Channel with a Cognitive Relay (IFC-CR) consists of a classical two-user interference channel in which the two independent messages are also {\em non-causally known} at a {\em cognitive relay} node. In this work a special class of IFC-CRs in which the sources do not create interference at the non-intended destinations is analyzed.
This special model results in a channel with two non-interfering point-to-point channels whose transmission is aided by an in-band cognitive relay, which is thus referred to as
the Parallel Channel with a Cognitive Relay (PC-CR). We determine the capacity of the PC-CR channel  to within \gap bits/s/Hz for all channel parameters. In particular, we present several new outer bounds which we achieve to within a constant gap by proper selection of Gaussian input distributions in a simple rate-splitting and superposition coding-based inner bound. The inner and outer bounds are numerically evaluated to show that the actual gap can be far less than \gap bits/s/Hz.
\end{abstract}


\section{Introduction}

The concepts of interference, cognition -- or non-causal message knowledge at a subset of network nodes -- and relaying have all been of great recent interest.
While capacities of the interference, cognitive and relay channels remain open for the general discrete memoryless channel models, they are known for certain classes of channels, and known to within constant gaps for the Gaussian noise counterparts.  In this work we focus on a particular channel model which illustrates the power of cognition and simultaneous relaying which we term the Parallel Channel with a Cognitive Relay (PC-CR), which encompasses several multi-user and cognitive/cooperative channel models, such as the Broadcast Channel (BC) and the Cognitive InterFerence Channel (CIFC).

\begin{figure}
\centering
\includegraphics[width=6cm]{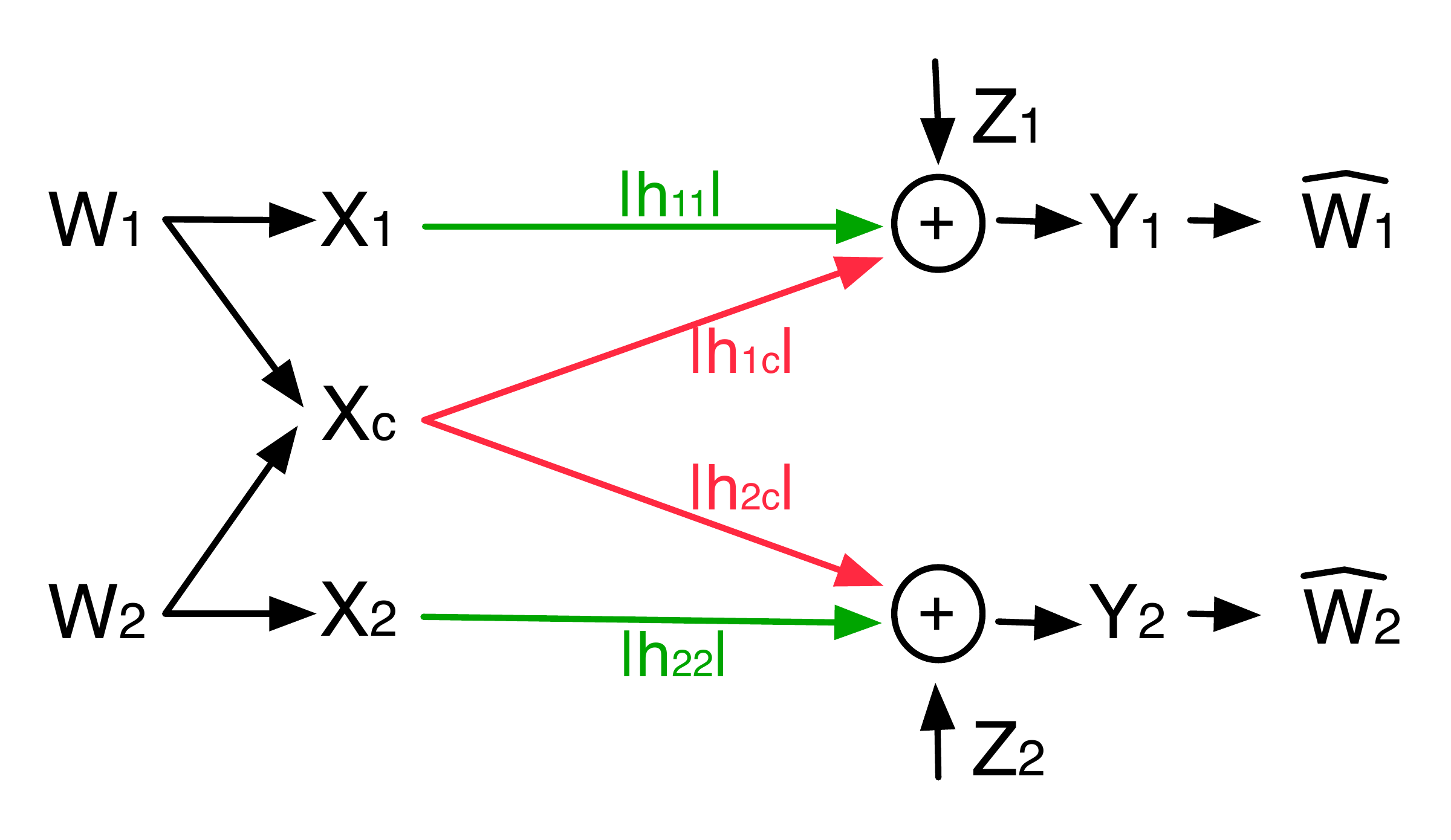}
\vspace{-0.2in}
\caption{The GPC-CR channel with two messages $W_1$ and $W_2$ known non-causally at the Cognitive Relay (CR).}\label{fig:GeneralModel}
\end{figure}

The PC-CR is a sub-channel of the more general interference channel with a cognitive relay (IFC-CR). The IFC-CR consists of a classical two-user interference channel in which the two independent messages, each known at the corresponding source node, are also {\em non-causally known} at a third, in-band transmitter node, which we term the {\em cognitive relay}.  This five-node channel generalizes a number of known channels including the BC, the IFC, and the CIFC. The PC-CR consists of an IFC-CR in which the two sources do not interfere at the non-intended destinations (see Fig.~\ref{fig:GeneralModel}) thus resulting into two parallel point-to-point channels whose communication is aided by a single cognitive relay which knows both messages non-causally.  As such, the emphasis is placed on the transmission strategy of the single cognitive relay: it may choose to help one source, the other, or both and it is this tradeoff that the PC-CR seeks to characterize.

\smallskip
{\bf Past Work.} The PC-CR has not been explicitly considered aside from the authors' previous work \cite{rini2010dublin}, 
where an outer bound for the general discrete memoryless IFC-CR was presented and then tightened for certain deterministic channels.
We then demonstrated that this outer bound is capacity for the high-SNR linear deterministic approximation of the Gaussian PC-CR.
In this work, we go beyond our previous work in \cite{rini2010dublin} by considering the Gaussian PC-CR model for finite SNR.

More generally, as the PC-CR is a subset of the IFC-CR, the literature for the latter channel is of direct relevance. The IFC with a relay was first introduced in \cite{Sahin_2007_1} and \cite{Sahin_2007_2}, where the message knowledge at the relay was obtained causally and non-causally, respectively.  Here we focus on the {\em non-causal} version of the problem~\cite{Sahin_2007_2}, also termed the ``broadcast channel with cognitive relays''~\cite{Jiang_BCCR}, and thus we omit the body of work related to the causal (i.e., non-cognitive) relay model.
In~\cite{Sahin_2007_2}, dirty-paper coding, beamforming and interference reduction techniques are combined in the derivation of an achievable rate region for the Gaussian single-input single-output (SISO) IFC-CR.
In~\cite{sridharan2008capacity}, the achievable region of~\cite{Sahin_2007_2} is further improved upon and a sum-rate outer bound based on the MIMO Gaussian C-IFC is proposed.
In~\cite{Jiang_BCCR}, an achievable rate region that contains all previously known regions is proposed.

\smallskip
{\bf Contributions and Paper Organization.} In this work we study the Gaussian PC-CR (GPC-CR).
We expand upon the limited prior work on the PC-CR, never explicitly considered in Gaussian noise, by:

\noindent $\bullet$ specializing the outer bound in \cite{rini2010dublin} to the GPC-CR (Section~\ref{sec:outerold})
and deriving two new outer bounds (Section~\ref{sec:outernew}), 

\noindent $\bullet$ specializing the achievable scheme in \cite{RTD-IFCCRstrongInterference} to the GPC-CR (Section~\ref{sec:inner}),

\noindent $\bullet$ determining capacity to within \gap bits/s/Hz by selecting Gaussian inputs in the achievable scheme that mimic the capacity achieving scheme of the corresponding high-SNR linear deterministic approximation channel \cite{rini2010dublin} (Section~\ref{sec:constant}),

\noindent $\bullet$ numerically evaluating inner and outer bounds, which indicate an even small than \gap bits/s/Hz gap
(Section~\ref{sec:constant}).

Section~\ref{sec:Conclusion and Future Work} concludes the paper.


\section{Channel Model}
\label{sec:Channel Model}
We use definitions and notation as in \cite{ThomasCoverBook},
and let $\Ccal(x) \defeq \log_2(1+x)$,
and $\alb \defeq 1-\al$ for $\al \ \in \ [0,1]$.
A complex-valued GPC-CR in {\em standard form} is a five-node network
(see Fig. \ref{fig:GeneralModel}) with
inputs ($X_1, X_2, X_c$) and outputs  ($Y_1, Y_2$) related through:
\begin{subequations}
\begin{align}
Y_1 &= |h_{11}| X_1+ |h_{1c}| X_c + Z_1,\\
Y_2 &= |h_{22}| X_2+ |h_{2c}| X_c + Z_2,
\end{align}
\label{eq:standard form channel definition}
\end{subequations}
where $Z_i \sim \Nc(0,1)$ and where the inputs are subject to a power constraint
$\E[|X_i|^2]\leq 1$, $i=\{1,2,c\}$.  The channel links $h_{i}$, $i \in \{11,22,1c,2c\}$
can be taken to be real-valued without loss of generality because receivers and transmitters
can compensate for the phase of the output signals.
The correlation among
the noises is irrelevant because the capacity of the channel without
receiver cooperation only depends on the noise marginal distributions~\cite{sato1978outer}.

A rate-pair $(R_1,R_2)$ is achievable if there exist
a sequence of encoding functions over $N$ channel uses
$X_i^N = X_i^N(W_i)$, $i\in\{1,2\}$,  and
$X_c^N = X_c^N(W_1, W_2)$ (where we note the non-causal message knowledge at the relay),
for messages $W_1$ and $W_2$ independent and uniformly distributed on
$[1:2^{N R_1}]$ and $[1:2^{N R_2}]$, respectively,
and a sequence of decoding functions
$\widehat{W}_i  = \widehat{W}_i(Y_i^N)$, $i\in\{1,2\},$
such that $\Pr\lsb  \widehat{W}_i \neq W_i  \rsb \to 0$ as $N\to\infty$ for $i\in\{1,2\}$.
The capacity region is the convex closure of the set of achievable
$(R_1,R_2)$-pairs.

\section{Outer Bounds}
\label{sec:outer}

In  this section we derive a series of outer bounds to the capacity region of the GPC-CR.
Our first outer bound is obtained by evaluating the outer bound we derived in \cite[Th. III.1]{rini2010dublin} for a general memoryless IFC-CR (which, as in Sato's outer bound for the BC \cite{sato1978outer}, exploits the fact that the capacity region only depends on the conditional marginal distributions of the channel) for the GPC-CR model.
%
The resulting outer bound is the tightest known for the GPC-CR but it is expressed as a function of four correlation parameters, which make its analytical manipulation difficult.
%
For this reason, we then proceed to derive a simpler (expression-wise) piecewise linear approximation of our first outer bound that will be used to derive the constant gap result in the ``large SNR'' regime.
%
Then, by following the approach of \cite[Th. III.7]{RTDjournal2}, we derive other two outer bounds by transforming the GPC-CR into a channel for which
tight bounds are available.
These two bounds 
will be used in the derivation of the constant gap results
in the ``small SNR'' regime.

\subsection{Tightest known outer bound}
\label{sec:outerold}

\begin{thm} 
\label{thm:outer bound I}
The capacity of a GPC-CR   
is contained in the region:
{\small
\begin{subequations}
\begin{align}
R_1 &\leq \Ccal( |h_{11}|^2+|h_{1c}|^2(1-|\rho_{2c}|^2) + 2 \Re \{ \rho_{1c}\} |h_{1c}| |h_{11} | ),
\label{eq:general outer bound R1} \\
R_2 &\leq \Ccal( |h_{22}|^2+|h_{2c}|^2(1-|\rho_{1c}|^2) + 2 \Re \{ \rho_{2c}\} |h_{2c}| |h_{22} | ),
\label{eq:general outer bound R2} \\
R_1 +R_2 & \leq \Ccal( |h_{22}|^2+|h_{2c}|^2+ 2 \Re \{ \rho_{2c}\}  |h_{2c}|   |h_{22}|)
 + \min_{\rho_z:|\rho_z| \leq 1}
\nonumber \\
&\hspace{-1.3cm}  \Ccal   \lb  |h_{11}|^2+ \f {|h_{1c}|^2}{|h_{2c}|^2}-2  \f {\Re\{\rho_z\} |h_{1c}|}{|h_{2c}|}-\f{\labs \rho_{1c} |h_{11}| | h_{2c}|+\rho_z- \f {|h_{1c}|}{|h_{2c}|}\rabs ^2} {|h_{2c}|^2 (1-|\rho_{2c}|^2)+1 }\rb
\nonumber \\
&-  \log(1-|\rho_z|^2)
\label{eq:general outer bound R1+R2 I} \\
R_1 +R_2 & \leq \Ccal( |h_{11}|^2+|h_{1c}|^2+ 2  \Re \{\rho_{1c}\} | h_{1c}||h_{11} | )
 + \min_{\rho_z:|\rho_z| \leq 1}
\nonumber \\
&\hspace{-1.3cm} \Ccal  \lb  |h_{22}|^2+ \f {|h_{2c}|^2}{|h_{1c}|^2}-2 \f {\Re \{\rho_z\}  |h_{2c}| } { |h_{1c}| }  -\f{ \labs \rho_{2c} |h_{22}| |h_{1c}| +\rho_z- \f {|h_{2c}|}{ |h_{1c}|}\rabs^2} {|h_{1c}|^2 (1-|\rho_{1c}|^2)+1 }\rb
\nonumber \\
&-  \log(1-|\rho_z|^2)
\label{eq:general outer bound R1+R2 II}
\end{align}
\label{eq:general outer bound}
\end{subequations}
}
for all $(|\rho_{1c}|, |\rho_{2c}|)\in[0,1]^2$ such that $|\rho_{1c}|^2+|\rho_{2c}|^2 \leq 1$.
\end{thm}
\begin{IEEEproof}
Consider the outer bound in \cite[Th. III.1]{rini2010dublin} for a general IFC-CR,  given by:
\begin{align*}
R_1    &\leq I(Y_1;X_1,X_c|X_2,Q),\\
R_2    &\leq I(Y_2;X_2,X_c|X_1,Q),\\
R_1+R_2&\leq I(Y_2;X_1,X_2,X_c,Q)+I(Y_1;X_1,X_c|Y_2,X_2,Q)\\
R_1+R_2&\leq I(Y_1;X_1,X_2,X_c,Q)+I(Y_2;X_2,X_c|Y_1,X_1,Q),
\end{align*}
for some input distribution that factors as $P_{Q,X_1,X_2,X_c}=P_{Q}P_{X_1|Q}P_{X_2|Q}P_{X_c|X_1,X_2,Q}$.
By specializing the above bound 
to the GPC-CR, the ``Gaussian maximizes entropy'' theorem \cite{ThomasCoverBook}
guarantees that the following Gaussian input:
\begin{align}
\begin{pmatrix}
X_1 \\ X_2 \\ X_c \\
\end{pmatrix}
\sim \Nc\lb \zerov,
\begin{pmatrix}
1                        &                   0  & \rho_{1c} \\
0                        &                   1  & \rho_{2c} \\
\rho_{1c}^*              &  \rho_{2c}^*         & 1 \\
\end{pmatrix}
\rb
\label{eq:jgin}
\end{align}
exhausts the outer region 
(because every mutual information term 
contains {\em all} the inputs).
The covariance matrix in~\reff{eq:jgin} is
positive semi-definite if and only if  
$|\rho_{1c}|^2+|\rho_{2c}|^2\leq 1$.
Evaluating the outer bound region 
for the jointly Gaussian input in~\reff{eq:jgin}
yields the region in~\reff{eq:general outer bound}.
The minimization over $\rho_z = \E[Z_1 Z_2^*]$ in each of the sum-rate bounds
is possible because the joint conditional distribution can be chosen so as to tighten the bound.
\end{IEEEproof}

Given that our outer bound in Th.~\ref{thm:outer bound I}
is expressed as the union over all feasible correlation coefficients
that satisfy $|\rho_{1c}|^2+|\rho_{2c}|^2= 1$,
it is not immediately useful in the derivation of the constant gap result.
To address this, we loosen the outer bound in Th.~\ref{thm:outer bound I}
and in doing so obtain a new, simpler piecewise linear outer bound expression,
which will be used in constant gap result in Section \ref{sec:constant}.

\begin{thm} {\bf Piecewise linear outer bound.}
\label{thm:linear outer bound}
The region in Th.~\ref{thm:outer bound I} is included into:
\begin{subequations}
\begin{align}
R_1 & \leq \Ccal (\max\{|h_{11}|^2,|h_{1c}|^2\}) + \log(4),
\label{eq:picewiselin R1}\\
R_2 & \leq \Ccal (\max\{|h_{22}|^2,|h_{2c}|^2\}) + \log(4),
\label{eq:picewiselin R2}\\
R_1+R_2 & \leq   \Ccal (\max\{|h_{22}|^2,|h_{2c}|^2\})
\nonumber\\& + \Ccal \lb \max\left\{|h_{11}|^2,\f {|h_{1c}|^2}{|h_{2c}|^2}\right\} \rb + \log(8),
\label{eq:picewiselin sum rate 1}\\
R_1+R_2 & \leq  \Ccal (\max\{|h_{11}|^2,|h_{1c}|^2\})
\nonumber\\& + \Ccal \lb \max\left\{|h_{22}|^2,\f {|h_{2c}|^2}{|h_{1c}|^2}\right\} \rb + \log(8).
\label{eq:picewiselin sum rate 2}
\end{align}
\label{eq:picewiselin}
\end{subequations}
\end{thm}
\begin{IEEEproof}
For the $R_1$-bound (and similarly for $R_2$-bound):
\begin{align*}
R_1
  &\leq \Ccal \lb  |h_{11}|^2+|h_{1c}|^2(1-|\rho_{2c}|^2)+2  \Re \{\rho_{1c}\}|h_{1c}| |h_{11}|  \rb
\\&\leq \Ccal \lb   (|h_{11}|+|h_{1c}|)^2  \} \rb
   \leq \Ccal \lb   4 \max\{|h_{11}|^2,  |h_{1c}|^2  \} \rb
\\&\leq \Ccal (\max\{|h_{11}|^2,|h_{1c}|^2\}) + \log(4).
\end{align*}
For the sum-rate bounds we let $\rho_z=0$ and we proceed as for the $R_1$-bound.
Note that  our gap of \gap bits/s/Hz will come form the term $\log(8)=\gap$~bits in~\reff{eq:picewiselin sum rate 1} and~\reff{eq:picewiselin sum rate 2} .
\end{IEEEproof}

\subsection{Outer bounds by transformation}
\label{sec:outernew}

We now present two additional outer bounds that are obtained by transforming the GPC-CR in the spirit of \cite[Th. II.7]{RTDjournal1}.
In particular, we show that the capacity of the PC-CR is contained into:
1) the capacity region of
a cognitive IFC and a point-to-point (P2P) channel in parallel, and
2) the capacity region of two P2P channels and a BC channel all in parallel.
The proofs can be found in the Appendix
and basically follows by showing that independent coding across the transformed parallel channels is optimal.

\begin{figure}
\centering
\includegraphics[width=6cm]{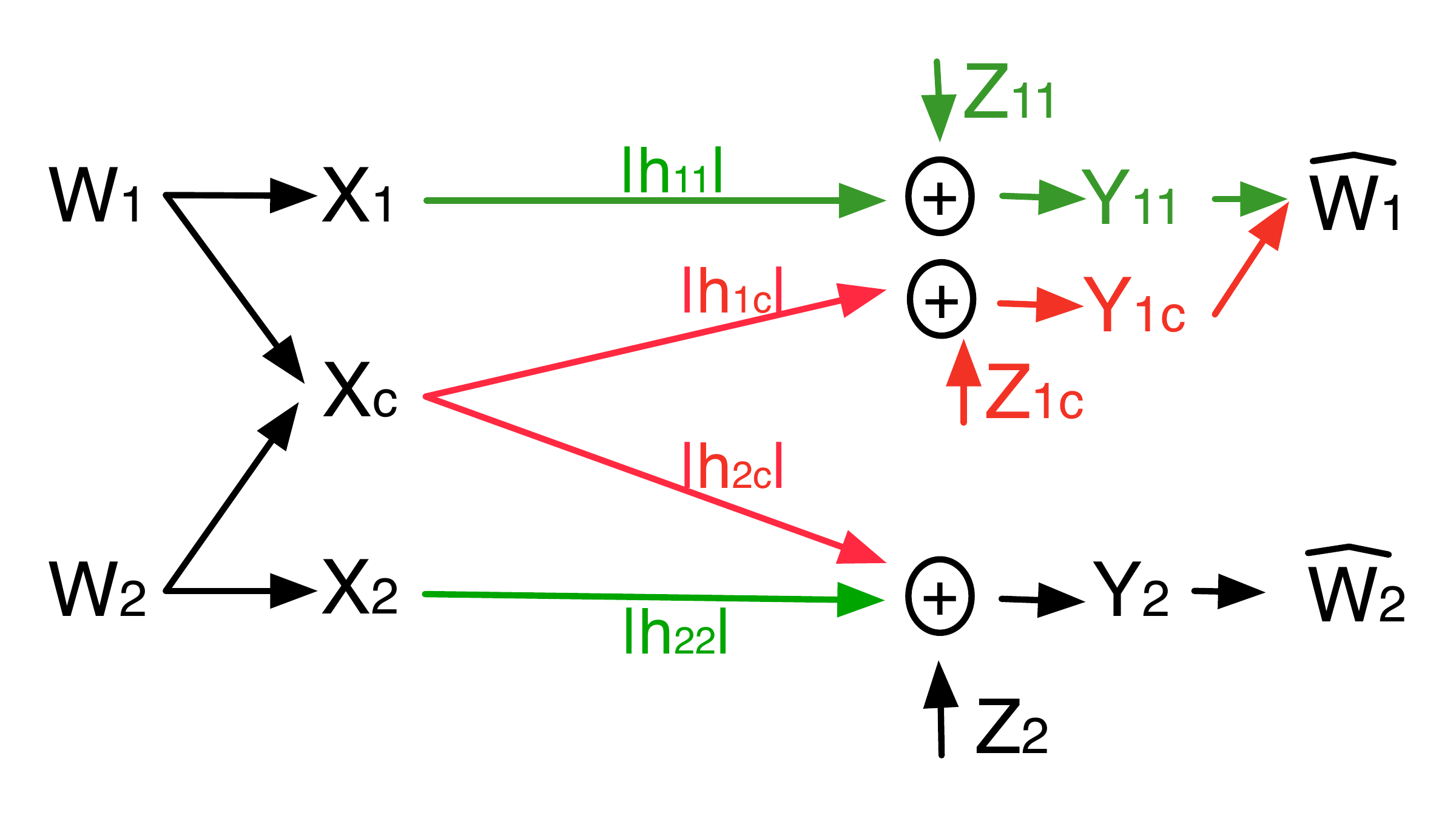}
\vspace{-0.2in}
\caption{The transformed channel for Th. \ref{th:CIFC+P2P outer bound}.}
\label{fig:cifc+p2p}
\includegraphics[width=6cm]{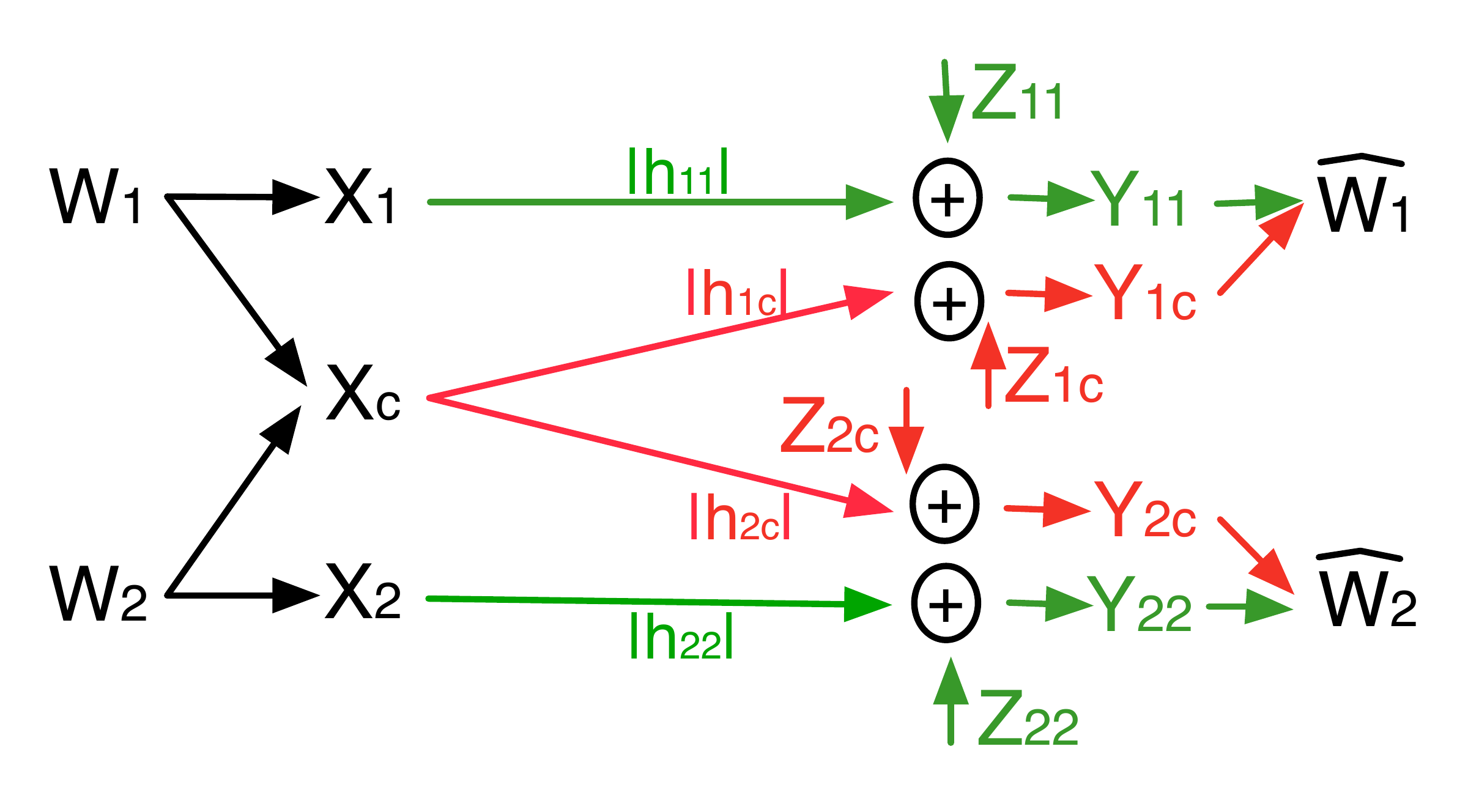}
\vspace{-0.2in}
\caption{The transformed channel for Th. \ref{th:P2P+BC+P2P outer bound}.}
\label{fig:bc+p2p+p2p}
\end{figure}

\begin{thm} {\bf CIFC+P2P outer bound.}
\label{th:CIFC+P2P outer bound}
The capacity of the  GPC-CR
is contained into the outer bound to the capacity region of the channel in Fig.~\ref{fig:cifc+p2p} given by:
\eas{
R_1 & \leq  \Ccal \lb \f{\al |h_{1c}|^2}{1-\sigma_{11}^2}\rb + \Ccal \lb  \f{|h_{11}|^2}{\sigma_{11}^2}  \rb ,  \\
R_1+R_2 &\leq  \Ccal \lb \alb  |h_{2c}|^2+ |h_{22}|^2+2 \sqrt{ \alb  } |h_{2c}| |h_{22}| \rb \\
 +& \lsb \Ccal \lb \f{\al |h_{1c}|^2 }{1-\sigma_{11}^2}\rb -\Ccal \lb  \al  |h_{2c}|^2  \rb \rsb^+ +  \Ccal \lb  \f{|h_{11}|^2}{\sigma_{11}^2}  \rb ,
}{\label{eq:CIFC+P2P outer bound}}
taken over the union of all  $\al \in [0,1]$, for any $\sigma_{11}^2\leq 1$.
\end{thm}

\begin{thm} {\bf Parallel P2P+BC outer bound.}
\label{th:P2P+BC+P2P outer bound}
The capacity of the GPC-CR
is contained into the capacity region of the channel in Fig.~\ref{fig:bc+p2p+p2p} given by:
\eas{
R_1 &\leq \Ccal \lb  \f{\alb |h_{1c}|^2 }{1-\sigma_{11}^2} \rb +\Ccal \lb \f{|h_{11}|^2} {\sigma_{11}^2} \rb,  \\
R_2 &\leq \Ccal \lb \f {\al  |h_{2c}|^2 } { \alb  |h_{2c}|^2 +1-\sigma_{22}^2}  \rb+\Ccal \lb \f{|h_{22}|^2}{\sigma_{22}^2} \rb,
}{\label{eq:P2P+BC+P2P outer bound}}
union over all $\al \in [0,1]$, for any $\sgs_1,\sgs_2 \leq 1$ such that
\ea{
\f{|h_{1c}|}{1-\sigma_{11}^2} \geq \f{|h_{2c}|}{1-\sigma_{22}^2}.
\label{eq:transformed channel degraded condition}
}
%
\end{thm}

\section{Inner Bound}
\label{sec:inner}
We now present a simple inner bound for the GPC-CR based on rate splitting and superposition coding. Somewhat surprisingly, we will show that this simple region achieves to within \gap bits/s/Hz of the outer bounds derived in the previous section.  The two sources split their message into common and private (sub)messages as in the classical Han and Kobayashi scheme for the IFC \cite{Han_Kobayashi81},  while the cognitive relay aids the two sources by relaying each of the (sub)messages.
%
%
Although this inner bound may be cast as a special case of the inner bound in \cite{jiang-achievable-BCCR}, it may alternatively be directly derived from the Han and Kobayashi region for the IFC \cite{Han_Kobayashi81} as follows.

\begin{thm}
\label{th:inner bound}
An achievable rate region for the GPC-CR  is the set of all non-negative $(R_1,R_2)$-pairs satisfying:
\begin{align*}
R_1 & \leq  I(Y_1; X_1 , U_{1c}| U_{2c},Q)  \nonumber \\
R_2 & \leq  I(Y_2; X_2 , U_{2c}| U_{1c},Q)  \nonumber \\
R_1+R_2 & \leq  I(Y_1; X_1,U_{1c},U_{2c}|Q)+I(Y_2; X_2| U_{1c},U_{2c},Q) \nonumber  \\
R_1+R_2 & \leq  I(Y_2; X_2,U_{2c},U_{1c}|Q)+I(Y_1; X_1| U_{1c},U_{2c},Q)  \nonumber \\
R_1+R_2 & \leq  I(Y_1; X_1,U_{2c}|U_{1c},Q)+ I(Y_2; X_2,U_{1c}|U_{2c},Q)  \nonumber  \\
2 R_1 +R_2 & \leq  I(Y_1; X_1,U_{2c},U_{1c}|Q)+I(Y_1;X_1| U_{1c},U_{2c},Q) \nonumber \\
& +I(Y_2; X_2,U_{1c}| U_{2c},Q)  \nonumber  \\
   R_1 +2 R_2 & \leq I(Y_2; X_2,U_{1c},U_{2c}|Q)+I(Y_2;X_2| U_{1c},U_{2c},Q) \nonumber \\
   & +I(Y_1; X_1,U_{2c}| U_{1c},Q).  \nonumber
\end{align*}
over the set of input distributions that factorize as
$P_{Q}P_{U_{1c},X_1|Q}P_{U_{2c},X_1|Q}P_{X_c|U_{1c},U_{2c},X_1,X_2,Q}$.
\end{thm}

We will consider the following Gaussian input in Th. \ref{th:inner bound}:
\begin{align*}
&U_i  \sim \Nc (0, 1) \ \text{independent  for all } i \in \{1p,2p,1c,2c\} \\
&X_1  =  \sqrt{\al_1}  U_{1p}+ \sqrt{\alb_1} U_{1c} , \text{ for } \al_1\in [0,1], \\
&X_2  =  \sqrt{\al_2}  U_{2p}+ \sqrt{\alb_2} U_{2c} , \text{ for } \al_2\in [0,1], \\
&X_c  =  \sqrt{\be_{1p}} U_{1p}+ \sqrt{\be_{1c}} U_{1c}
       + \sqrt{\be_{2p}} U_{2c}+ \sqrt{\be_{2c}} U_{2c},
 \\& \text{ for }(\be_{1p},\be_{1c},\be_{2p},\be_{2c})\in[0,1]^4:
       \be_{1p}+\be_{1c}+\be_{2p}+\be_{2c}= 1,
\end{align*}
the resulting rate bounds are omitted for brevity, but can be found in \cite{RTDjournal3}.
Note that with the above choice of inputs, the channel is effectively equivalent to a classical Gaussian IFC.

\section{Constant Gap}
\label{sec:constant}

In this section we show that the inner bound derived in Section \ref{sec:inner} lies to within a constant gap of the outer bounds derived in Section \ref{sec:outer}.  To prove this result we take inspiration from the capacity achieving scheme for the high-SNR linear deterministic approximation of the GPC-CR we derived in~\cite{rini2010dublin} (which we do not repeat this here due to space constraints).
We do however note that in deriving the capacity result in~\cite{rini2010dublin}, different achievability schemes were needed for different parameter regimes.  In deriving the constant gap result for the Gaussian Pc-CR, we will directly mimic, or choose, the inputs as the Gaussian analogy to the high-SNR capacity-achieving scheme in~\cite{rini2010dublin}. In fact, the different subcases shown in Table~\ref{table:gaussian} directly parallel the regimes for the high-SNR deterministic model, as further elaborated upon in \cite{RTDjournal3}.

\begin{table*}
\caption{The various PC-CR channel parameter regions, each of which employes different achievability scheme.}
\label{table:gaussian}
\centerline{
\begin{tabular}{|l|l|l|l|l|}
\hline
Label  & Case & Subcase & Outer Bound & Achievability scheme (absence implies zero rate)  \\
\hline
$\p{ \rm Weak\\\rm (W)}$
   & $\p{ \vspace{-.25cm} \\
         |h_{11}|^2 \geq |h_{1c}|^2,\\
         |h_{22}|^2 \geq |h_{2c}|^2,\\
         \vspace{-.25cm} \\}$
   &
   & $\p{ R_1 \leq \Ccal( |h_{11}|^2)+2, \\
          R_2 \leq \Ccal( |h_{22}|^2)+2, }$
   & $X_1 = U_{1p}, X_2=U_{2p}$ \\
\hline
$\p{ \rm Strong\\\rm (S)}$
   & $\p{|h_{11}|^2 < |h_{1c}|^2,\\
         |h_{22}|^2 < |h_{2c}|^2, }$
   & $|h_{11}|^2 \geq\f {|h_{1c}|^2} {|h_{2c}|^2}$
   &
   $\p{   \vspace{-.25cm} \\
   R_1 & \leq \Ccal ( |h_{1c}|^2 )+2\\
         R_2 & \leq \Ccal (|h_{2c}|^2 )+2\\
         R_1 + R_2 & \leq \Ccal (|h_{2c}|^2) + \Ccal \lb |h_{11}|^2 \rb +3}$
   &  S.1: $\begin{array}{l}
\mbox{Point A:} X_1=U_{1p}, X_c = U_{1c}, X_2=U_{2p}  \\
\mbox{Point B:} \p{X_1=U_{1p}, X_2=U_{2p}\\X_c= \be_{2c}U_{2c}+ \be_{2p}U_{2p}} \end{array}$ \\
\cline{3-5}
   &
   & $|h_{11}|^2 <\f {|h_{1c}|^2} {|h_{2c}|^2}$
   & $\p{
   \vspace{-.25cm} \\
   R_1 & \leq \Ccal (|h_{1c}|^2 )+2\\
         R_2 & \leq \Ccal (|h_{2c}|^2 )+2\\
         R_1 + R_2 & \leq \Ccal (|h_{2c}|^2)+ \Ccal \lb \f {|h_{1c}|^2} {|h_{2c}|^2} \rb+3 }$
   & S.2 : $\begin{array}{l}
\mbox{ Point A:} X_1 =U_{1p}, X_c=X_1 \\
\mbox{ Point B:} X_c =\beta_{1c}U_{1c}+\beta_{2p}U_{2p}
 \end{array}$ \\
\hline
$\p{ \rm Mixed\\\rm (M)}$
    & $\p{|h_{11}|^2< |h_{1c}|^2,\\
          |h_{22}|^2\geq  |h_{2c}|^2,} $
    & $|h_{11}|^2 \geq\f {|h_{1c}|^2} {|h_{2c}|^2}$
    & $\p{\vspace{-.25cm} \\
     R_1 &\leq \Ccal ( |h_{1c}|^2) +2\\
           R_2 &\leq \Ccal ( |h_{22}|^2) +2\\
           R_1+R_2 & \leq \Ccal ( |h_{22}|^2 )+\Ccal( |h_{11}|^2)+3}$
    & M.1: $\begin{array}{l}
\mbox{Point A:} X_1=U_{1p},X_c = U_{1c} \\
\mbox{Point B:} X_1=U_{1p}, X_2=U_{2p}\end{array}$ \\
\cline{3-5}
    &
    & $|h_{11}|^2 <\f {|h_{1c}|^2} {|h_{2c}|^2}$
    & $\p{ \vspace{-.25cm} \\
    R_1 &\leq \Ccal ( |h_{1c}|^2)+2 \\
           R_2 &\leq \Ccal ( |h_{22}|^2)+2 \\
           R_1+R_2 &\leq \Ccal (  |h_{22}|^2 )+\Ccal \lb  \f {|h_{1c}|^2} {|h_{2c}|^2}\rb+3}$
    & M.2 : $\begin{array}{l}
\mbox{Point A:} X_c = U_{1c}, X_2=U_{2p} \\
\mbox{Point B:} X_c = \beta_{1p}U_{1p}+\beta_{2c}U_{2c}, X_2=U_{2c}\end{array}$ \\
\hline
\end{tabular}
}
\end{table*}

\begin{thm}
\label{thm:constant gap}
The inner bound of Th.~\ref{th:inner bound} achieves capacity to within \gap bit/sec/Hz.
\end{thm}

\begin{figure}
\centering
\includegraphics[width=9.5 cm]{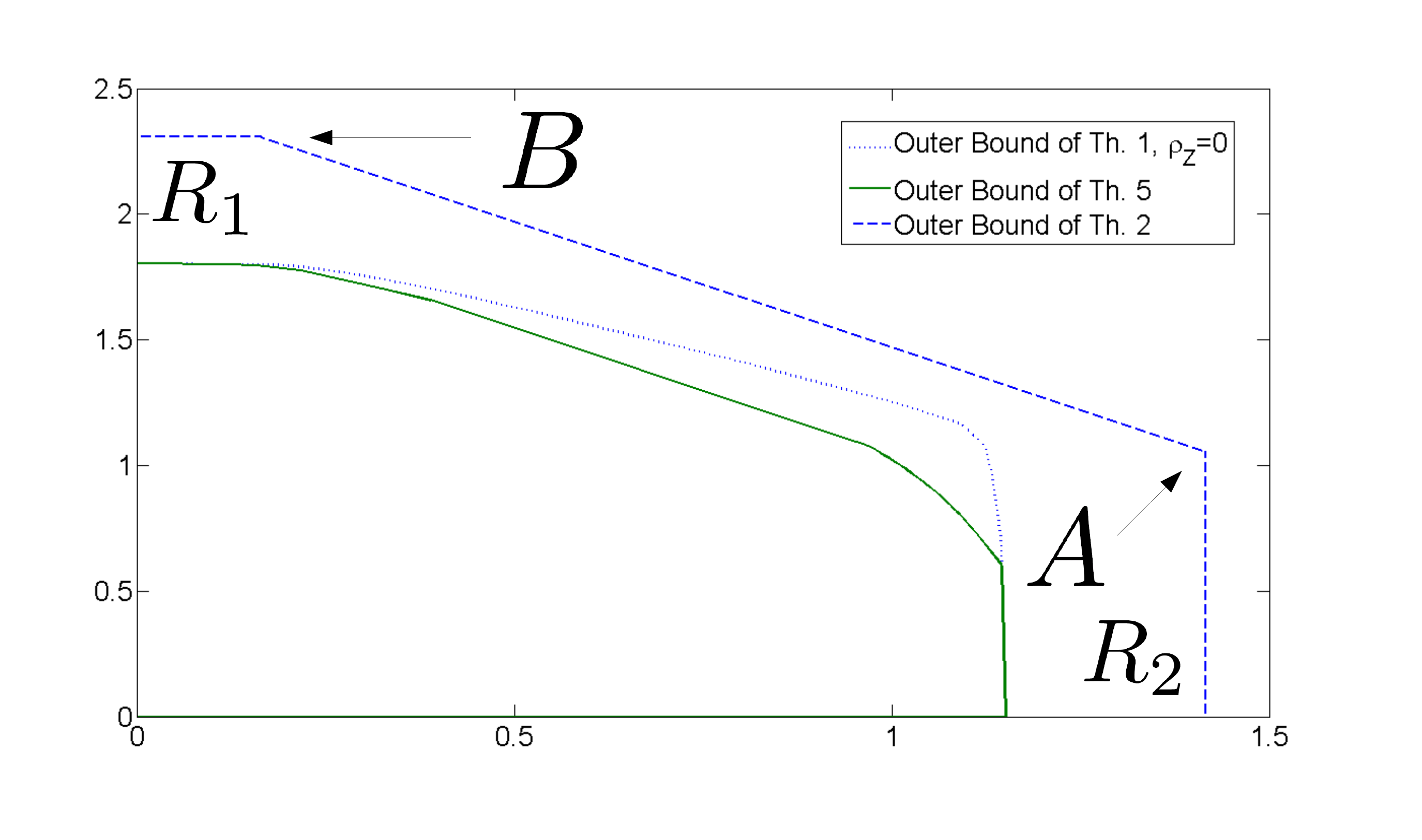}
\vspace*{-0.2in}
\caption{A comparison of the outer bounds of Th. \ref{thm:outer bound I} and Th. \ref{thm:linear outer bound} and the inner bound of Th. \ref{th:inner bound} for the channel with $h_{11}=2 h_{1c}=h_{2c}=h_{22}/5=1$.} 
\label{fig:InnerBoundAndOuterBound}
\end{figure}

\begin{IEEEproof}
To establish this constant gap result we take inspiration
from the proof of the capacity of the high-SNR linear deterministic PC-CR in
\cite[Cor. IV.2]{rini2010dublin} and we partition the parameter space into three regions,
each of which uses an achievability scheme inspired by the deterministic counterpart in the
corresponding regime.
The full and lengthy proof is provided in \cite{RTDjournal3}, we outline some of the key ideas next.
The ``piece-wise linear'' outer bound region in \reff{eq:picewiselin}
\footnote{
For ``large SNR''
, i.e. $\min_{i\in\{11,1c,2c,22\}}\{|h_{i}|^2\}\geq 1$,
we compare the inner bound with the outer bound in Th.~\ref{thm:linear outer bound}.
For ``small SNR'' the same proof applies, but instead of the Th.~\ref{thm:linear outer bound}
one must use Th. \ref{th:CIFC+P2P outer bound} and Th. \ref{th:P2P+BC+P2P outer bound} (see \cite{RTDjournal3}).}
has two Pareto optimal corner points (see Fig. \ref{fig:InnerBoundAndOuterBound}):
\eas{
A = \Big( \reff{eq:picewiselin R1}, \min\big\{\reff{eq:picewiselin sum rate 1}-\reff{eq:picewiselin R1},\reff{eq:picewiselin sum rate 2}-\reff{eq:picewiselin R1},\reff{eq:picewiselin R2}\big\} \Big),
\label{eq:corner point A}
\\
B = \Big( \min\big\{\reff{eq:picewiselin sum rate 1}-\reff{eq:picewiselin R2},\reff{eq:picewiselin sum rate 2}-\reff{eq:picewiselin R2},\reff{eq:picewiselin R1}\big\},\reff{eq:picewiselin R2} \Big).
\label{eq:corner point B}
}{\label{eq:corner point A and B}}
For each parameter regime we show the achievability of the two corner points A and B;
because of the min-expressions in~\reff{eq:corner point A and B} we will have to consider
different sub-cases.
In the following we assume without loss of generality that~\reff{eq:picewiselin sum rate 2} dominates~\reff{eq:picewiselin sum rate 1}. Table~\ref{table:gaussian} summarizes the different schemes used to achieve
a constant gap from the corner points in \reff{eq:corner point A and B} in the right column
When/why  each of these schemes are useful may be found in the Appendix and \cite{RTDjournal3} in the
following the different sub-cases are briefly discussed.

\paragraph{Case W: ``weak cognition at both decoders'' regime}
When $|h_{11}|^2\geq |h_{1c}|^2$ and $|h_{22}|^2\geq |h_{2c}|^2$, that is,
when the power of the signal from the source to its destination
is stronger than the power of the signal from the relay to the destination,
the corner points A and B in \reff{eq:corner point A and B} coincide.
%
By leaving the relay silent (i.e., $\be_{1p}=\be_{1c}=\be_{2p}=\be_{2c}=0$)
and transmitting $X_i=U_{ip}$ (i.e., $\al_i=1$)
we can achieve $R_i\leq \Cc(|h_{11}|^2)$, $i\in\{1,2\}$,
which is to within two bits
the outer bound in Th. \ref{thm:linear outer bound}.
In this regime there is no critical advantage in using the cognitive relay
to relay the message of either source because the cognitive relay is received
at small power.

\paragraph{Case S:  ``strong cognition at both decoders'' regime}
If $|h_{11}|^2<|h_{1c}|^2$ and $|h_{22}|^2< |h_{2c}|^2$,
the gain of the link from the cognitive relay to each destination is larger
than the direct link from the source to the destination.
%
%
Consider achieving corner point A in~\reff{eq:corner point A}
(point B  may be achieved in an similar way by reversing the role of the users):
here the cognitive relay cooperates with source~1 in sending a common message and with source~2 in sending  a private message.
Since $|h_{22}|^2< |h_{2c}|^2$, the common message of source~1 can be decoded at destination~2 without any rate penalty.
The power allocated to the private message of source~2 by the cognitive relay is such
that the interference it creates  at destination~1 is at or below the noise floor,
as in \cite{etkin2008Gaussian} for the classical IFC.
This choice of transmit powers causes a gap of at most one bit
for the achievable rate $R_1$ from point A on the outer bound,
but results in a gap of three bits for the achievable rate $R_2$.
%

%
\paragraph{Case M: ``mixed cognition'' regime}
%
When $|h_{11}|^2 \leq |h_{1c}|^2$ and $|h_{22}|^2 \geq |h_{2c}|^2$  the direct link between source~1 and destination~1 has a smaller gain than the link from the cognitive relay to destination~1, while the opposite holds for source~2.
%
In trying to achieve corner point A in~\reff{eq:corner point A}, both transmitters send a private message to the intended destination while the cognitive relay relays sends a common message for source~1.
Decoding  the common message of source~1 allows destination~2 to strip the interference from its channel output.
In trying to achieve corner  B in~\reff{eq:corner point B}, the cognitive relay employs a strategy  similar to the ``strong cognition at both decoders'' regime: %
it cooperates with the both sources simultaneously and relays the private message of source~1 with such amplitude as  to cause an interference at destination~2 at or below the noise floor.
This specific choice of power levels causes a loss of performance of one bit for rate $R_2$ but significantly boosts the achievable rate $R_1$.
%
\end{IEEEproof}

%
While we have analytically shown that the achievable rate region of Th. \ref{th:inner bound} achieves to within \gap bits/s/Hz the outer bounds in Th.~\ref{thm:linear outer bound},
Fig.~\ref{fig:InnerBoundAndOuterBound} shows, at least for certain channel parameters, that the gap
from the less analytically tractable outer bound of  Th. \ref{thm:outer bound I} is much less.

\section{Conclusion and Future Work}
\label{sec:Conclusion and Future Work}
While the exact capacity of the GPC-CR is still unknown, we demonstrated inner and outer bounds which 
lie to within \gap bits/s/Hz from each other. This gap result was arrived at through insights gained from the capacity achieving schemes for the high-SNR linear deterministic approximation of the GPC-CR.   We  view this channel as a building block toward the more general IFC-CR in which interfering links between the two non-cognitive users are present. The PC-CR has allowed us to focus on the role of cognitive cooperation first, before incorporating interference created by the non-intended transmitter, which is the subject of ongoing work.

\section*{Acknowledgment}
The work of the authors was partially funded by NSF under awards number 0643954 and 1017436. The contents of this article are solely the responsibility of the authors and do not necessarily represent the official views of the NSF.

\bibliographystyle{IEEEtran}
\bibliography{steBib1}

\appendices
\newpage

\section{Proof of Th.\ref{th:CIFC+P2P outer bound}}
\label{app:CIFC+P2P outer bound}

The capacity of the PC-CR in Fig.~\ref{fig:GeneralModel}
is contained into the capacity of the transformed channel
given in Fig. \ref{fig:cifc+p2p}
where the input output  relationship is described by:
\begin{subequations}
\begin{align}
\lsb \p{Y_{11} \\ Y_{1c}}\rsb &= \lsb \p{ \; |h_{11} |X_1+ Z_{11}\\ |h_{1c}| X_c+ Z_{1c} \\ }\rsb \\
Y_2 &= |h_{22}| X_2 + |h_{2c}| X_c,
\end{align}
\label{eq: transformed channel CIFC+P2P}
\end{subequations}
where $Z_{11}\sim \Nc \lb 0,\sigma_{11}^2 \rb$, $\sigma_{11}^2\in[0,1]$,
independent of $Z_{1c}\sim\Nc \lb 0,1-\sigma_{11}^2 \rb$, where the inputs
are subject to $\E[|X_i|^2] \leq 1$, $i \in \{1,c,2\}$. Since  $Y_1 \sim Y_{11}+Y_{1c}$,
the capacity of the transformed channel is no smaller than the capacity of the original channel.
By Fano's inequality:
\begin{subequations}
\begin{align}
&N( R_1 - \epsilon_N) \nonumber
    \leq  I( Y_{11}^N, Y_{1c}^N; W_1)
\\&   =   I( Y_{11}^N; W_1 )+I(Y_{1c}^N; W_1 |Y_{11}^N)       \nonumber
\\& \leq  I( Y_{11}^N; W_1 )+I(Y_{1c}^N; W_1 |Y_{11}^N, W_2)  \nonumber
\\& =     h( Y_{11}^N )
         -h( Y_{11}^N| W_1, X_1^N ) \nonumber
\\&      +h( Y_{1c}^N|Y_{11}^N, W_2, X_2^N)
         -h( Y_{1c}^N|Y_{11}^N, W_2, X_2^N, W_1, X_1^N, X_c^N)  \nonumber
\\& \leq  h( Y_{11}^N )
         -h( Y_{11}^N| X_1^N ) \nonumber
\\&      +h( Y_{1c}^N|X_2^N)
         -h( Y_{1c}^N|X_2^N,X_c^N)  \label{eq:wsxcde}
\\& =     I( Y_{11}^N; X_1^N )+I(Y_{1c}^N;X_c^N|X_2^N)  \nonumber
\\           & \leq  N \big( I(Y_{11};X_1|Q) + I( Y_{1c}; X_c|X_2, Q) \big),  \nonumber
\end{align}
\label{eq:CIFC+P2P}
\end{subequations}
where in~\reff{eq:wsxcde} we have used ``conditioning reduces entropy'' to drop $(Y_{11}^N, W_2)$
in $h( Y_{1c}^N|Y_{11}^N, W_2, X_2^N)$ and the fact that $Y_{1c}^N$ is independent of everything else
once conditioned on $X_c^N$.

For the sum-rate, by Fano's inequality:
\begin{subequations}
\begin{align}
&N( R_1 + R_2- 2\epsilon_N)\nonumber
\\&  \leq  I( Y_{11}^N, Y_{1c}^N; W_1) + I(Y_2^N; W_2) \nonumber
\\&  \leq  I( Y_{11}^N, Y_{1c}^N, Y_{2}^N, W_2; W_1) + I(Y_2^N; W_2) \nonumber
\\& = h(Y_2^N) - h(Y_2^N| W_1, W_2)   \nonumber
\\& + I( Y_{11}^N; W_1| Y_{2}^N, W_2)
    + I( Y_{1c}^N; W_1| Y_{2}^N, W_2, Y_{11}^N) \nonumber
\\& \leq
      I(Y_2^N; X_2^N,X_c^N)  \nonumber
\\& + I( Y_{11}^N; X_1^N)
    + I( Y_{1c}^N; X_c^N| Y_{2}^N, X_2^N) \label{edcvfr}
\\&= N\big( I(Y_2; X_2,X_c|Q)+ I( Y_{11}; X_1|Q)
    + I( Y_{1c}; X_c| Y_{2}, X_2, Q) \big) \nonumber
\end{align}
\label{eq:CIFC+P2P R1+R2}
\end{subequations}
where \reff{edcvfr} follows from steps similar to those used to derive the $R_1$-bound.
The outer bound is obtained as the union over all the distributions that factors
$P_{Q,X_1,X_2,X_c}=P_{Q}P_{X_1|Q}P_{X_2|Q}P_{X_c|X_1,X_2,Q}$.
Moreover, by the ''Gaussian maximizes entropy property" \cite{ThomasCoverBook}, we have that zero-mean complex Gaussian inputs maximize all the bounds in \reff{eq:CIFC+P2P outer bound} and thus we let
\begin{align}
\begin{pmatrix}
X_1 \\ X_2 \\ X_c \\
\end{pmatrix}
\sim \Nc\lb \zerov,
\begin{pmatrix}
1                        &                   0  & 0 \\
0                        &                   1  & \alb \\
0                        &  \alb^*         & 1 \\
\end{pmatrix}
\rb
\label{eq:jgin 1}.
\end{align}
Note that we can fix $\cov(X_1,X_2)=0$ wlg since none of the rate bounds depends on the correlation of these two RVs. 
The expression in \reff{eq:CIFC+P2P outer bound} is finally obtained by maximizing the sum rate bound of \reff{eq:CIFC+P2P R1+R2}, with the inputs parametrization of \reff{eq:jgin 1}, over the joint distribution of $P_{Y_1,Y_2|X_1,X_2,X_c}$ for fixed marginals $P_{Y_1|X_1,X_2,X_c}$ and $P_{Y_2,Y_2|X_1,X_2,X_c}$ as in \cite{RTDjournal2}.


\section{Proof of Th. \ref{th:P2P+BC+P2P outer bound}}
\label{app:P2P+BC+P2P outer bound}

As in the proof of Th. \ref{th:CIFC+P2P outer bound},
the capacity of the PC-CR in Fig.~\ref{fig:GeneralModel}
is contained into the capacity of the transformed channel
given in Fig.~\ref{fig:bc+p2p+p2p}
where the input output  relationship is described by:
\begin{subequations}
\begin{align}
\lsb \p{Y_{11} \\ Y_{1c} }\rsb &=\lsb \p { |h_{11}|X_1+ Z_{11} \\ |h_{1c}|  X_c+ Z_{1c} } \rsb ,\\
\lsb \p{Y_{22} \\ Y_{2c} }\rsb &=\lsb \p { |h_{22}|X_2+ Z_{22}  \\ |h_{2c}| X_c+ Z_{2c} } \rsb ,
\end{align}
\label{eq: transformed channel P2P+BC+P2P}
\end{subequations}
for  $Z_{11}\sim \Nc(0,\sigma_{11}^2)$, $\sigma_{11}^2\in[0,1]$, independent of $Z_{1c}\sim \Nc(0,1-\sigma_{11}^2)$, and
     $Z_{22}\sim \Nc(0,\sigma_{22}^2)$, $\sigma_{22}^2\in[0,1]$ independent of $Z_{2c}\sim \Nc(0,1-\sigma_{22}^2)$,
and where the inputs are subject to $\E[|X_i|^2] \leq 1$, $i \in \{1,c,2\}$. Note the transformed
channel in Fig.~\ref{fig:bc+p2p+p2p} consists of two point-to-point channels and a BC all in parallel.
The original channel in Fig.~\ref{fig:GeneralModel} is a degraded version of the
channel in Fig.~\ref{fig:bc+p2p+p2p} since
$Y_i \sim Y_{ii}+Y_{ic}$, $i\in\{1,2\}$.
As in the proof of Th. \ref{th:CIFC+P2P outer bound},  we first show that independent coding
across the parallel channels is optimal and then obtain the capacity region of the channel as
the sum of the capacity of parallel channels.
By Fano's inequality:
\begin{subequations}
\begin{align}
&N( R_1 - \epsilon_N) \leq  I( Y_{11}^N, Y_{1c}^N; W_1) \nonumber
\\&  \leq   I( Y_{1c}^N; W_1 )+I(Y_{11}^N; W_1)  \label{eq:P2P+BC+P2P 1}
\\&  \leq   N \lb I( Y_{1c}; X_c |U )+I(Y_{11}; X_1) \rb \label{eq:P2P+BC+P2P 2}
\end{align}
\label{eq:P2P+BC+P2P R1}
\end{subequations}
where \reff{eq:P2P+BC+P2P 1}  follows from the same steps as  \reff{eq:CIFC+P2P},
 and  \reff{eq:P2P+BC+P2P 2}  from the rate bound for the degraded BC  in \cite{gallager1974capacity},
where the auxiliary random variable $U$ forms the Markov chain $U-X_c-Y_1-Y_2$ under the condition \reff{eq:transformed channel degraded condition}.
Similarly we have that
\begin{subequations}
\begin{align}
&N( R_2 - \epsilon_N)  \leq  I( Y_{22}^N, Y_{2c}^N; W_1) \nonumber
\\&  \leq   I( Y_{2c}^N; W_1 )+I(Y_{22}^N; W_1)  \label{eq:P2P+BC+P2P 3}
\\&  \leq   N \lb I( Y_{2c}; U )+I(Y_{22}; X_1) \rb \label{eq:P2P+BC+P2P 4}
\end{align}
\label{eq:P2P+BC+P2P R2}
\end{subequations}
where \reff{eq:P2P+BC+P2P 3}  follows from the same steps as  \reff{eq:CIFC+P2P} for user~2,
 and  \reff{eq:P2P+BC+P2P 4}  from the rate bound for the degraded BC in \cite{gallager1974capacity} for the same $U$ as in \reff{eq:P2P+BC+P2P R1}.
The bounds  of \reff{eq:P2P+BC+P2P R1} and \reff{eq:P2P+BC+P2P R2} do not depend on the joint distribution of $P_{X_1,X_2,X_c}$ but only on the marginal distribution of each RV. For this reason we can take $X_c \perp X_1$ and $X_c \perp X_2$  wlg while $X_{1} \perp X_2$ by definition.
From this consideration and the ``Gaussian maximizes entropy" property of \cite{ThomasCoverBook},  it follows that the optimal distribution of $X_{11}$ and $X_{22}$  is zero-mean complex Gaussian while from the ``entropy power inequality" of \cite{gallager1974capacity} it follows that the optimal $U$ and $X_c$ must be proper complex Gaussian.
The expression of \reff{eq:P2P+BC+P2P outer bound} is obtained by noting that bounds are maximized when the power constraint is met with equality.

\end{document}